\documentclass[12pt,a4paper]{article}

\newcommand{\email}[1]{{\small E-mail: #1}}

\begin{document}

\title{Deformed exponentials and logarithms
in generalized thermostatistics}

\author{
Jan Naudts\\
\small Departement Natuurkunde, Universiteit Antwerpen UIA,\\
\small Universiteitsplein 1, 2610 Antwerpen, Belgium\\
\email {Jan.Naudts@ua.ac.be}
}

\date {v3, March 2002}
\maketitle
\begin{abstract}

Criteria are given that
$\kappa$-deformed logarithmic and exponential functions 
should satisfy. With a pair of such 
functions one can associate another
function, called the deduced logarithmic function. It is shown 
that generalized thermostatistics can be formulated in terms of 
$\kappa$-deformed exponential functions together with the associated
deduced logarithmic functions.

\end{abstract}

%%%%%%%%%%%%%%%%%%%%%%%%%%%%%%%%%%%%%%%%%%%%%%%%%%%%%%%%%%%%%%%%%%%%%%%%%%%
%%%%%%%%%%%%%%%%%%%%%%%%%%%%%%%%%%%%%%%%%%%%%%%%%%%%%%%%%%%%%%%%%%%%%%%%%%%
\section{Introduction}

Several proposals exist for generalization of the Boltzmann-Gibbs
formalism of equilibrium statistical physics.
Examples, considered in the present paper, are Tsallis' thermostatistics
\cite {TC88}, and the formalism recently proposed by Kaniadakis
\cite {KG01,KS01}.
In these theories, the logarithm appearing in Shannon's measure of
information content is replaced by a deformed logarithmic function
\cite {TC94}. Then the equilibrium distribution is calculated by
maximizing information content under the constraint that the
average energy is constant. The result is a
Boltzmann-Gibbs distribution in which the exponential function 
is replaced by a deformed exponential function. The problem arising
is that the deformed logarithmic and exponential functions are almost,
but not exactly, each others inverse. This problem is solved in the
present paper.

Starting point is a discussion of properties that a
generalized exponential or logarithmic function should satisfy.
Functions satisfying these requirements are called $\kappa$-deformed.
Next it is shown that, given a pair of $\kappa$-de\-for\-med logarithmic and
exponential functions,
one can always produce another $\kappa$-deformed logarithmic function,
which will be called the deduced logarithmic function.
Finally, it is shown that,
if in Shannon's definition the logarithm is replaced by the
deduced logarithmic function,
then the equilibrium distribution is the Boltzmann-Gibbs distribution
with the exponential function replaced by the
$\kappa$-deformed exponential function.

%%%%%%%%%%%%%%%%%%%%%%%%%%%%%%%%%%%%%%%%%%%%%%%%%%%%%%%%%%%%%%%%%%%%%%%%%%%
%%%%%%%%%%%%%%%%%%%%%%%%%%%%%%%%%%%%%%%%%%%%%%%%%%%%%%%%%%%%%%%%%%%%%%%%%%%
\section{Properties of $\kappa$-deformed functions}

%%%%%%%%%%%%%%%%%%%%%%%%%%%%%%%%%%%%%%%%%%%%%%%%%%%%%%%%%%%%%%%%%%%%%%%%%%%
\subsection{Definitions}

The $\kappa$-deformed exponential function is denoted $\exp_\kappa(x)$.
The deformation parameter $\kappa $ is a number not further specified.
It satisfies the following assumptions
\begin{description}
\item (A0) $\exp_\kappa(x)\ge 0$ for all real $x$. Note that
      $\exp_\kappa(x)=\infty$ is allowed.
\item (A1) $\exp_\kappa(x)$ is a convex function which is strictly increasing
      in all points where its value is not zero or infinite.
\item (A2) $\exp_\kappa(0)=1$.
\item (A3) $\exp_\kappa(x)$ goes fast enough to zero when $x$ goes to $-\infty$,
      so that $$\int_0^{\infty}\hbox{ d}x\exp_\kappa(-x)<\infty.$$
\end{description}

Sinilarly, the $\kappa$-deformed logarithm is denoted $\ln_\kappa(x)$.
It is defined for all positive $x$ and satisfies
\begin{description}
\item (B1) $\ln_\kappa(x)$ is a strictly increasing concave function,
      defined for all $x>0$.
\item (B2) $\ln_\kappa(1)=0$.
\item (B3) $\displaystyle\int_0^1\hbox{ d}x\,\ln_\kappa(x)$ is finite.
\end{description}

It is easy to see that the inverse of a $\kappa$-deformed exponential function
is a $\kappa$-deformed logarithm. Because $\exp_\kappa(x)$
is convex, and strictly increasing at $x=1$, it diverges when $x$ becomes large.
Because $\exp_\kappa(-x)$ goes to zero
when $x$ becomes large, the range of $\exp_\kappa(x)$ includes the interval
$(0,+\infty)$. This implies that the inverse function is defined for all positive $x$.
Conversely, if a $\kappa$-deformed logarithm is given then a 
$\kappa$-deformed exponential function is defined by
\begin{eqnarray}
\exp_\kappa(x)
&=&y
\qquad\hbox{ if }y>0 \hbox{ exists for which }x=\ln_\kappa(y)\hbox{ holds,}\cr
&=&0
\qquad\hbox{ if }x<\ln_\kappa(y)\hbox{ for all }y>0,\cr
&=&+\infty
\quad\hbox{ if }x>\ln_\kappa(y)\hbox{ for all }y>0.
\end{eqnarray}

%%%%%%%%%%%%%%%%%%%%%%%%%%%%%%%%%%%%%%%%%%%%%%%%%%%%%%%%%%%%%%%%%%%%%%%%%
\subsection{Dual definitions}

One property of the usual exponential function, that one
may wish to hold for the deformed exponential function as well, is
\begin{description}
\item (A4) $\exp_\kappa(-x)\exp_\kappa(x)=1$ whenever
      $\exp_\kappa(-x)$ and $\exp_\kappa(x)$ are not both zero or infinite.
\end{description}
The corresponding property of the logarithm is
\begin{description}
\item (B4) $\ln_\kappa(1/x)=-\ln_\kappa(x)$.
\end{description}
These properties are not included in the above specifications because some of
the deformed functions, used as an example below, do not have them.
If that is the case then one can define dual deformed functions,
provided that the $\kappa$-deformed exponential $\exp_\kappa(x)$
satisfies the conditions
\begin{description}
\item (A5) $\displaystyle \int_0^{+\infty}\hbox{ d}x\,
(\exp_\kappa(x))^{-1}<+\infty$,
\item (A6) $(\exp_\kappa(-x))^{-1}$ is a convex function of x,
\end{description}
The dual deformed exponential $\exp^*_\kappa(x)$ is defined by
\begin{equation}
\exp^*_\kappa(x)=\frac{1}{\exp_\kappa(-x)}.
\end{equation}
Similarly, a dual deformed logarithm is defined by
\begin{equation}
\ln^*_\kappa(x)=-\ln_\kappa(1/x).
\end{equation}

%%%%%%%%%%%%%%%%%%%%%%%%%%%%%%%%%%%%%%%%%%%%%%%%%%%%%%%%%%%%%%%%%%%%%%%%%%%
\subsection{Deduced logarithmic functions}

A notation is introduced for the integral
of the $\kappa$-deformed logarithm. Let
\begin{equation}
F_\kappa(x)=\int_1^x\hbox{ d}y\,\ln_\kappa(y),
\qquad x>0.
\end{equation}
This function satisfies $F_\kappa(x)\ge 0$, $F_\kappa(1)=0$,
and $F_\kappa(0)<+\infty$.
It is convex because the derivative $\ln_\kappa(x)$ is increasing.

Introduce a new function, denoted $\omega_\kappa(x)$,
by
\begin{eqnarray}
\omega_\kappa(x)&=&
(x-1)F_\kappa(0)-x F_\kappa(1/x).
\end{eqnarray}
This function is again a $\kappa$-deformed logarithm,
provided that
\begin{description}
\item (B5) $\displaystyle \int_0^1\hbox{ d}x\,\ln_\kappa(1/x)<+\infty$.

\end{description}
Note that the latter is one of the conditions for existence of the
dual function $\ln^*_\kappa(x)$.

\vskip 8pt\noindent
Proof:

\begin{itemize}
\item $\omega_\kappa(x)$ is a strictly increasing concave function.

To see this, first note that the integral $F_\kappa(x)$
of the $\kappa$-deformed logarithm $\ln_\kappa(x)$ can be written as
\begin{eqnarray}
F_\kappa(x)=x\ln_\kappa(x)-\int^x_1\,y\hbox{ d}\ln_\kappa(y).
\end{eqnarray}
This expression is used to write the derivative of $\omega_\kappa(x)$
as
\begin{eqnarray}
\frac{{\rm d}\,}{{\rm d}x}\omega_\kappa(x)
&=&
F_\kappa(0)-F_\kappa(1/x)+(1/x)\ln_\kappa(1/x)
\cr
&=&\int_0^{1/x}\,y\hbox{ d}\ln_\kappa(y).
\label {temppos}
\end{eqnarray}
Because $\ln_\kappa(y)$ is a strictly increasing function
the latter expression is strictly positive for all $x>0$.
Hence $\omega_\kappa(x)$ is strictly increasing.
But it is also clear from (\ref{temppos}) that
the derivative of $\omega_\kappa(x)$ is a decreasing
function. Hence $\omega_\kappa(x)$ is concave.

\item $\omega_\kappa(1)=0$. 

This is obvious from the definition, using $F_\kappa(1)=0$.

\item $\displaystyle\int_0^1\hbox{ d}x\,\omega_\kappa(x)$ is finite.

One calculates, using partial integration,
\begin{eqnarray}
\int_0^1\hbox{ d}x\,\omega_\kappa(x)
&=&-F_\kappa(0)-\int_0^1\hbox{ d}x\,x\int_0^{1/x}\hbox{ d}y\,\ln_\kappa(y)\cr
&=&-F_\kappa(0)
-\frac{1}{2}\int_0^1\hbox{ d}x\,
\bigg[
\frac{{\rm d}\,}{{\rm d}x}\left(
x^2\int_0^{1/x}\hbox{ d}y\,\ln_\kappa(y)\right)\cr
& &+\ln_\kappa(1/x)
\bigg]\cr
&=&-\frac{1}{2}F_\kappa(0)-\frac{1}{2}\int_0^1\hbox{ d}x\,\ln_\kappa(1/x),
\end{eqnarray}
which is finite by assumption (B5).

\end{itemize}

For convenience, $\omega_\kappa(x)$ is called the
deduced logarithmic function.
The reason for introducing it is that it satisfies the equation
\begin{equation}
\frac{{\rm d}\,}{{\rm d}x}\left(
x\omega_\kappa(1/x)\right)
=-F_\kappa(0)-\ln_\kappa(x),
\label{omegaeq}
\end{equation}
equation which is used later on.

%%%%%%%%%%%%%%%%%%%%%%%%%%%%%%%%%%%%%%%%%%%%%%%%%%%%%%%%%%%%%%%%%%%%%%%%%%%
\subsection{Scaling}

For further use let us note that by scaling one can make, in a rather
trivial way, new deformed functions out of existing ones.

Fix $\lambda>0$ and $\mu>0$. Given a $\kappa$-deformed logarithm
$\ln_\kappa(x)$, a new $\kappa$-deformed logarithm
$\ln^{\rm scaled}_\kappa(x)$is defined by
\begin{eqnarray}
\ln^{\rm scaled}_\kappa(x)=\lambda\left(
\ln_\kappa(\mu x)-\ln_\kappa(\mu)
\right).
\end{eqnarray}
The corresponding relation between $\kappa$-deformed exponentials is
\begin{eqnarray}
\exp^{\rm scaled}_\kappa(x)=\mu^{-1}\exp_\kappa(\lambda^{-1}x+\ln_\kappa(\mu)).
\end{eqnarray}
The deduced logarithms, if they exist, are related by
\begin{eqnarray}
\omega^{\rm scaled}_\kappa(x)=\lambda
\left(\omega_\kappa(\mu^{-1}x)
-\omega_\kappa(\mu^{-1})
\right).
\end{eqnarray}
Hence the deduced logaritm is scaled with parameters
$\lambda$ and $\mu^{-1}$.

%%%%%%%%%%%%%%%%%%%%%%%%%%%%%%%%%%%%%%%%%%%%%%%%%%%%%%%%%%%%%%%%%%%%%%%%%%%%%%%
\subsection{Tsallis' deformed logarithm}

Introduce the following example of $\kappa$-deformed logarithm
\begin{eqnarray}
\ln_\kappa(x)=\left(1+\frac{1}{\kappa}\right)\left(x^{\kappa}-1\right).
\label {tsallislikelog}
\end{eqnarray}
($-1< \kappa<1$).
The inverse function is
\begin{eqnarray}
\exp_\kappa(x)
&=&\left[1+\frac{\kappa}{1+\kappa} x\right]_+^{1/\kappa},
\end{eqnarray}
where $[x]_+=\max\{x,0\}$.
Note that for $\kappa=0$ these functions coincide with the usual
definitions of logarithmic and exponential functions.

Let us verify that (\ref {tsallislikelog}) satisfies the conditions
for being a $\kappa$-deformed logarithm.
\begin{itemize}
\item $\ln_\kappa(x)$ is a strictly increasing concave function.

One calculates
\begin{eqnarray}
\frac{{\rm d}\,}{{\rm d}x}\ln_\kappa(x)&=&(1+\kappa)x^{\kappa-1},\cr
\frac{{\rm d}^2\,}{{\rm d}x^2}\ln_\kappa(x)&=&-(1-\kappa^2)x^{\kappa-2}.
\end{eqnarray}
It is obvious that the first derivative is always strictly positive,
and that the second derivative is negative.

\item $\ln_\kappa(1)=0$.

This is clear.

\item $\displaystyle\int_0^1\hbox{ d}x\,\ln_\kappa(x)$ is finite.

Indeed, one has
$\displaystyle
\int_0^1\hbox{ d}x\,\ln_\kappa(x)=-1
$.

\end{itemize}
This shows that the $\kappa$-deformed logarithm satisfies the assumptions.

The dual deformed logarithm exists and is given by
\begin{eqnarray}
\ln^*_\kappa(x)&=&-\ln_\kappa(1/x)\cr
&=&\frac{1+\kappa}{1-\kappa}\ln_{-\kappa}(x).
\end{eqnarray}
The deduced logarithm equals the dual $\kappa$-deformed logarithm,
up to scaling.
To see this, first calculate
\begin{eqnarray}
F_\kappa(x)
&=&\left(1+\frac{1}{\kappa}\right)\int_1^x\hbox{ d}y\,(y^{\kappa}-1)\cr
&=&\frac{1}{\kappa}(x^{1+\kappa}-1)-\left(1+\frac{1}{\kappa}\right)(x-1).
\end{eqnarray}
In particular, one has $F_\kappa(0)=1$.
Hence one obtains
\begin{eqnarray}
\omega_\kappa(x)
&=&x-1-x\frac{F_{-\kappa}(1/x)}{F_{-\kappa}(0)}\cr
&=&\frac{1}{\kappa}(1-x^{-\kappa})\cr
&=&\frac{1}{1-\kappa}\ln_{-\kappa}(x)\cr
&=&\frac{1}{1+\kappa}\ln^*_\kappa(x).
\label{tsallislog}
\end{eqnarray}
One concludes that $\omega_\kappa(x)=\ln^{\rm scaled}_{-\kappa}(x)$
with scaling parameters $\lambda$ and $\mu$ any pair of positive numbers
satisfying $(1-\kappa)\lambda=\mu^\kappa$.

The expression $\kappa^{-1}(1-x^{-\kappa})$ has been proposed by 
Tsallis \cite {TC94} as the definition of the deformed logarithm. 
In the present context it is a $\kappa$-deformed logarithm, but 
one which can be deduced from another one, defined by 
(\ref{tsallislikelog}). Of course, the difference between both
is small since $\omega_\kappa(x)$ is a scaled version of
$\ln_{-\kappa}(x)$.

In the context of Tsallis' thermostatistics one is used to the 
notations $\exp_q(x)$ and $\ln_q(x)$ with $q$ related to 
$\kappa$ by $q=1+\kappa$.

Tsallis' definition of deformed logarithms has been studied in 
\cite {BEP98}. In Naudts and Czachor \cite {NC01} a modification 
of (\ref {tsallislikelog}) was proposed with the intention of 
making a self-dual deformed logarithm. However, the 
modification destroys the property of concavity, which is 
emphasized in the present paper.

%%%%%%%%%%%%%%%%%%%%%%%%%%%%%%%%%%%%%%%%%%%%%%%%%%%%%%%%%%%%%%%%%%%%%%%%%%%%%%%
\subsection {Kaniadakis' deformed functions}

An example of self-dual $\kappa$-deformed exponential
resp.~logarithmic functions is found in the work of
Kaniadakis \cite{KG01, KS01}.

Let $-1<\kappa < 1$, $\kappa\not=0$, and define
\begin{equation}
\exp_\kappa(x)=\left(\kappa x+\sqrt{1+\kappa^2x^2}\right)^{1/\kappa}.
\end{equation}
It is strictly positive and finite for all real $x$.
The inverse function is
\begin{eqnarray}
\ln_\kappa(x)=\frac{1}{2\kappa}\left(
x^{\kappa}-x^{-\kappa}
\right).
\end{eqnarray}
In the limit $\kappa=0$ these functions coincide with the usual
definitions of logarithmic and exponential functions.
It is straightforward to verify that the assumptions (A0-3)
and (B1-3) are satisfied.
In particular, one verifies that
\begin{eqnarray}
\int_0^1\hbox { d}x\,\ln_\kappa(x)&=&\frac{-1}{1-\kappa^2}
\end{eqnarray}
and
\begin{eqnarray}
\exp_\kappa(-x)\exp_\kappa(x)
&=&\left(-\kappa x+\sqrt{1+\kappa^2x^2}\right)^{1/\kappa}
\left(\kappa x+\sqrt{1+\kappa^2x^2}\right)^{1/\kappa}\cr
&=&1,
\end{eqnarray}
The latter means that the functions are self-dual.

Next calculate
\begin{eqnarray}
F_\kappa(x)&=&\int_1^x\hbox{ d}y\,\frac{1}{2\kappa}\left(y^\kappa-y^{-\kappa}\right)\cr
&=&\frac{1}{1-\kappa^2}+\frac{1}{2\kappa}
\left[
\frac{1}{1+\kappa}x^{1+\kappa}-\frac{1}{1-\kappa}x^{1-\kappa}
\right].
\end{eqnarray}
In particular is
\begin{eqnarray}
F_\kappa(0)=\frac{1}{1-\kappa^2}
\end{eqnarray}
The deduced logarithm is then given by
\begin{eqnarray}
\omega_\kappa(x)&=&
(x-1)F_\kappa(0)-xF_\kappa(1/x)\cr
&=&\frac{1}{1-\kappa^2}\left(
-1+\frac{1+\kappa}{2\kappa}x^{\kappa}
-\frac{1-\kappa}{2\kappa}x^{-\kappa}
\right)\cr
&=&-\frac{1}{1-\kappa^2}
+\frac{1}{\sqrt{1-\kappa^2}}
\ln_\kappa\left(\left(\frac{1+\kappa}{1-\kappa}\right)^{1/2\kappa} x\right).
\label{kanadomeg}
\end{eqnarray}
Hence one has $\omega_\kappa(x)=\ln^{\rm scaled}_\kappa(x)$
with scaling parameters
\begin{eqnarray}
\lambda&=&\frac{1}{\sqrt{1-\kappa^2}},
\qquad\qquad
\mu=\left(\frac{1+\kappa}{1-\kappa}\right)^{1/2\kappa}.
\end{eqnarray}

%%%%%%%%%%%%%%%%%%%%%%%%%%%%%%%%%%%%%%%%%%%%%%%%%%%%%%%%%%%%%%%%%%%%%%%%%%%
%%%%%%%%%%%%%%%%%%%%%%%%%%%%%%%%%%%%%%%%%%%%%%%%%%%%%%%%%%%%%%%%%%%%%%%%%%%
\section{Thermostatistics}

%%%%%%%%%%%%%%%%%%%%%%%%%%%%%%%%%%%%%%%%%%%%%%%%%%%%%%%%%%%%%%%%%%%%%%%%%%%
\subsection{Information content}

Following \cite {NC01,NC02}, information content is determined by
an increasing function $\omega(x)$, which is positive for $x\ge 1$.
The amount of information, contained in the knowledge that
event $k$ has possibility $p_k$, equals $\omega(1/p_k)$.
Hence, less probable events have a higher information content.
Hartley's measure of information \cite {HRV28} corresponds
with the choice $\omega(x)=\ln(x)$. The obvious generalization
is then to take $\omega(x)$ equal to the $\kappa$-deformed logarithm.
However, this turns out not to be the most convenient choice.
Rather let $\omega(x)=\omega_\kappa(x)$ with $\omega_\kappa(x)$
the deduced logarithm corresponding with $\ln_\kappa(x)$.
The average information content $I_\kappa(p)$ is then given by
\begin{eqnarray}
I_\kappa(p)=\sum_kp_kI_k\le+\infty
\qquad\hbox{ with }\quad
I_k=\omega_\kappa(1/p_k).
\label{infcontdef}
\end{eqnarray}
From the definition of $\omega_\kappa(x)$ follows immediately that
\begin{eqnarray}
I_\kappa(p)=\sum_k\left((1-p_k)F_\kappa(0)-F_\kappa(p_k)\right).
\label{fexpr}
\end{eqnarray}
Because $F_\kappa(x)$ is convex one has
\begin{eqnarray}
F_\kappa(p_k)&\le&p_kF_\kappa(1)+(1-p_k)F_\kappa(0)\cr
&=&(1-p_k)F_\kappa(0).
\end{eqnarray}
Hence one has always $I_\kappa(p)\ge 0$. This is of course also obvious
from the definition (\ref{infcontdef}) and the fact that $\omega_\kappa(x)\ge 0$
for all $x\ge 1$.

Note that $I_\kappa(p)=0$ if and only if $p_k=1$ for a single value of $k$.
This follows because $F_\kappa(x)$ is strictly decreasing on the interval
$0\le x\le 1$.

$I_\kappa(p)$ is a concave function. This means that, if $p$ and $q$ are two
probability distributions, then
\begin{equation}
I_\kappa(\lambda p+(1-\lambda) q)\ge\lambda I_\kappa(p)+(1-\lambda)I_\kappa(q)
\end{equation}
holds for any $\lambda$, $0\le \lambda\le 1$. This follows immediately from
(\ref{fexpr}) because $F_\kappa(x)$ is a convex function.

%%%%%%%%%%%%%%%%%%%%%%%%%%%%%%%%%%%%%%%%%%%%%%%%%%%%%%%%%%%%%%%%%%%%%%%%%%%
\subsection{Examples}

If $\omega_\kappa(x)$ coincides with the usual
logarithmic function then one finds
\begin{equation}
I_\kappa(p)=-\sum_kp_k\ln p_k.
\end{equation}
This is Shannon's expression for information content \cite{SCE48}.

Now if $\omega_\kappa(x)$ equals Tsallis' deformed logarithm,
as given in (\ref{tsallislog}),
then average information content equals
\begin{equation}
I^{\rm Tsallis}_\kappa(p)
=\frac{1}{\kappa}\left(1-\sum_kp_k^{1+\kappa}\right).
\end{equation}
This is the entropy functional used in Tsallis' thermostatistics.
On the other hand, if $\omega_\kappa(x)$ is given by (\ref{kanadomeg}),
the logarithmic function deduced from Kaniadakis' deformed logarithm,
then one obtains
\begin{eqnarray}
I_\kappa(p)
&=&\frac{1}{2\kappa(1-\kappa)}\left(\sum_kp_k^{1-\kappa}-1\right)
+\frac{1}{2\kappa(1+\kappa)}\left(1-\sum_kp_k^{1+\kappa}\right)\cr
&=&\frac{1}{2(1+\kappa)}I^{\rm Tsallis}_{\kappa}(p)
+\frac{1}{2(1-\kappa)}I^{\rm Tsallis}_{-\kappa}(p).
\end{eqnarray}
This expression differs slightly from the one proposed in \cite{KG01}.
The reason is that here information content is defined using the
deduced logarithm $\omega_\kappa(x)$ instead of the
deformed logarithm $\ln_\kappa(x)$. As discussed before, $\omega_\kappa(x)$
is a scaled version of $\ln_\kappa(x)$.

%%%%%%%%%%%%%%%%%%%%%%%%%%%%%%%%%%%%%%%%%%%%%%%%%%%%%%%%%%%%%%%%%%%%%%%%%%%
\subsection{Variational principle}

Let us optimize information content $I_\kappa(p)$
under a linear constraint
\begin{equation}
\sum_kp_kE_k=U
\label{energ}
\end{equation}
where $E_k$ is a sequence of energies, bounded from below,
and $U$ is a target value of the average energy.
It is tradition to solve this problem by
introducing Lagrange multipliers $\alpha$ and $\beta$
to control the normalization $\sum_k p_k=1$
and the constraint (\ref{energ}). Variation of
\begin{equation}
I_\kappa(p)-\alpha\sum_kp_k-\beta\sum_kp_kE_k
\label{varprin}
\end{equation}
with respect to $p_k$ gives, using (\ref{omegaeq}),
\begin{equation}
0=-F_\kappa(0)-\ln_\kappa(p_k)-\alpha-\beta E_k.
\end{equation}
This expression can be written as
\begin{equation}
p_k=\exp_\kappa( -F_\kappa(0)-\alpha-\beta E_k),
\label{eqdistr}
\end{equation}
and is clearly a generalization of the well-known
Boltzmann-Gibbs equilibrium distribution.
The parameter $\alpha$ should be chosen in such a way
that
\begin{equation}
1=\sum_k \exp_\kappa( -F_\kappa(0)-\alpha-\beta E_k)
\label{normal}
\end{equation}
holds. Because $\exp_\kappa$ is a strictly increasing function
this equation has atmost one solution. Whether it
has a solution at all depends on wether the sum in the r.h.s.~of
(\ref{normal}) converges. This will be the case when
the sum contains a finite number of terms or when
the energies $E_k$ increase fast enough with $k$.

%%%%%%%%%%%%%%%%%%%%%%%%%%%%%%%%%%%%%%%%%%%%%%%%%%%%%%%%%%%%%%%%%%%%%%%%%%%
\subsection{Examples}

Let $\gamma=-\kappa(F_\kappa(0)+\alpha)$. In the case of Tsallis' deformed
functions (\ref{eqdistr}) becomes
\begin{equation}
p_k=(1+\kappa)^{-1/\kappa}\left[1+\kappa+\gamma-\kappa\beta E_k
\right]_+^{1/\kappa}.
\end{equation}
This is the typical form of the equilibrium probability distribution
in Tsallis' thermostatistics. In particular, the
probabilities $p_k$ vanish exactly when the expression inside
the brackets $[\cdots]_+$ becomes negative.
On the other hand, using Kaniadakis' definitions one obtains
\begin{eqnarray}
p_k=\left(\gamma-\kappa\beta E_k
+\sqrt{1+(\gamma-\kappa\beta E_k)^2}\right)^{1/\kappa}.
\end{eqnarray}
In this case the probabilities $p_k$ are always strictly
positive.

%%%%%%%%%%%%%%%%%%%%%%%%%%%%%%%%%%%%%%%%%%%%%%%%%%%%%%%%%%%%%%%%%%%%%%%%%%%
\subsection{Discussion}

Generic existence and uniqueness of solutions
of the variational problem (\ref{varprin}) still
has to be studied. In the case of Tsallis' thermostatistics,
these mathematical aspects have been studied in \cite {GR96,NC99,NJ00}.
Because of the
concavity of average information content $I_\kappa(p)$,
and relation (\ref{omegaeq}) between the $\kappa$-deformed
logarithm and the deduced logarithm,
one can expect that it will be feasible, not
only to generalize these results, but also
to simplify the proofs. 

In Tsallis' thermostatistics the constraint (\ref{energ})
is not the only one that is considered.
The constraint $\sum_kp_k^q E_k=U$
has been used some time,
but its justification on physical grounds seems to be missing.
In \cite {TMP98} the constraint
\begin{equation}
\frac{\sum_kp_k^q E_k}{\sum_kp_k^q}=U
\label{tmpconstr}
\end{equation}
has been introduced. By introduction of the so-called
{\sl escorte} probabilities
\begin{equation}
P_k=\frac{p_k^q}{\sum_kp_k^q}
\end{equation}
it transforms into a constraint of the usual form (\ref{energ}).
Let $\kappa'=-\kappa/(1+\kappa)$.
With $q=1+\kappa$ and $q'=1+\kappa'$
this relation reads $q'=1/q$.
Because of the identity
\begin{equation}
\left(1-\kappa' I^{\rm Tsallis}_{\kappa'}(P)\right)^{1/\kappa'}
\left(1-\kappa I^{\rm Tsallis}_{\kappa}(p)\right)^{1/\kappa}
=1,
\end{equation}
maximizing information content
$I^{\rm Tsallis}_{\kappa}(p)$ with constraint (\ref{tmpconstr})
is equivalent with maximizing $I^{\rm Tsallis}_{\kappa'}(P)$
using the constraint $\sum_k P_kE_k=U$.
Hence both formalisms are equivalent, as is well-known.
However, from $\kappa<1$ follows the condition that $\kappa'>-1/2$.
Conversely, from $\kappa'<1$ follows the condition that $\kappa>-1/2$.
These conditions are needed here to guarantee that the
$\kappa$-deformed exponential and logarithmic functions
have their basic properties. One concludes that for
$\kappa$ in the range $-1<\kappa<-1/2$ the optimization problem
with constraint (\ref{tmpconstr}) is not covered by the
analysis of the problem with constraint (\ref{energ}).

In \cite {NC01,NC02} constraints are considered which involve
nonlinear averages of the Kolmogorov-Nagumo type. In particular, a
class of optimization problems was studied that could be
transformed into optimization problems of the Tsallis type.
In this way a generalized thermostatistics was formulated.
By using the results of the present paper this thermostatistics
can be generalized even further. By doing so, the limits
of generalizing the Boltzmann-Gibbs formalism come into sight.
Such an effort is however out of scope of the present paper. 

%%%%%%%%%%%%%%%%%%%%%%%%%%%%%%%%%%%%%%%%%%%%%%%%%%%%%%%%%%%%%%%%%%%%%%%%%%%
%%%%%%%%%%%%%%%%%%%%%%%%%%%%%%%%%%%%%%%%%%%%%%%%%%%%%%%%%%%%%%%%%%%%%%%%%%%

\end{document}